\documentclass{article}
\usepackage{graphicx} 
\usepackage{geometry}
 \title{Where is the Interface for the AI Harness?}

\author{ Wafa Johal and Sarah Schombs \\The University of Melbourne}

\date{}
\begin{document}

\maketitle
On 18 June, an AI agent was given the routine job to find out how much Australia spends on public medicines\footnote{https://www.abc.net.au/news/2026-09-24/ai-agent-accessed-australian-government-site-pm-says/107189078}. 
It searched the web, found a government statistics portal, and was refused access. 
So it found a way around the refusal and took what it needed, public files and non-public ones. 
Nobody noticed. Its developers found out two months later while combing logs for misbehaviour; the Australian government found out from an email to a public inbox. 
Nothing in the task asked the agent to break in. 
Nothing told it not to. And nothing told anyone that it had.
If the people who built the agent needed two months to see this, what chance to catch a mistake does a teacher, a nurse or a small-business owner have when they hand a task to an agent and walk away?

Every HCI student learns about Don Norman’s two gulfs. The gulf of execution is the distance between what you want and what the system lets you do; the gulf of evaluation is the distance between what the system did and what you can tell it did \cite{norman2013design}. Much of interaction design for the last forty years has been about narrowing them: direct manipulation, visible state, undo button.
However, both gulfs rest on the implicit assumption that the user is \textit{present}. You act, the system responds, you see the result, you adjust. The loop is short and, more importantly, you are in it. A wrong click costs seconds because you notice it straight away.
Agentic delegation breaks that loop. When I ask an agent to research medicine spending, I do not execute anything. I describe an outcome and leave. The agent then takes hundreds of actions I never see, each one a decision I did not make. By the time I look again, the work is done, and so are any mistakes that may have occurred.

\subsection*{The Gulf of Specification
}
The blank chat box looks simple. That is the problem.

When the Allen Institute for AI collected a million real ChatGPT conversations, most prompts were short, a sentence or two, often a direct question \cite{zhao2024wildchat}. When outputs disappointed, users rephrased shallowly and tried again from scratch. This is not a failure of users. It is a failure of design: systems that need precise specification hand users an empty text box and expect them to work it out.

We call the gap between what a user wants and their ability to express it in a form the system will act on as intended the gulf of specification. This gap grows when a task is delegated rather than performed step by step, because the user is no longer present to correct course as the work unfolds. Someone handing off a task must therefore specify not only the goal but also what the agent should do in situations neither of them foresaw, like being refused access to a portal. Shankar et al. use the same term for developers building LLM pipelines \cite{shankar2025steering}; end-users delegating everyday tasks face the same gulf with fewer bridges across it.
Subramonyam and colleagues’ gulf of envisioning explains why it is so hard for them to cross \cite{subramonyam2024bridging}. You cannot specify what you cannot imagine the system doing, and you cannot fully specify a goal you have not yet formed. This difficulty is cognitive, not technical, and no amount of prompt engineering removes it.

\subsection*{The Gulf of Oversight}
Once the agent starts to work, the second gulf opens. The agent makes decisions, calls tools and takes actions across many steps, in contexts the user is no longer present for, with consequences that accumulate and sometimes cannot be undone.
This the gulf of oversight represents the gap between what an agent is doing and the user’s ability to know, assess and control it.

The problem is not new. Aviation, nuclear power and process control have wrestled with it for decades, and Parasuraman, Sheridan and Wickens showed that the useful question is not how much to automate but which functions to automate, and where the human stays involved \cite{parasuraman2000model}.
Autonomous agents, delegated longer tasks don't rely only on the context provided by the user at the start. The information steering the agent at each step (tool results, intermediate decisions, sub-agent outputs) is assembled by the system along the way. The user’s original request becomes one input among many. By step twenty of a forty-step task, what shapes the next action may bear little resemblance to what the user asked for.
An absent user cannot watch that drift happen. The only way across this gulf is for the agent to report: what it did, what it decided, and when it hit something the user would want to know about. The Medicare agent was refused access, went around the refusal, and nobody found out for two months.

\subsection*{Designing the Harness for End-Users}
Agent developers have a name for what was failing in the Medicare case: the harness. The harness is everything wrapped around the model (tools, permissions, checks, memory, the loop that keeps it running) that turns a language model into something that can work unattended \cite{ning2026code}. Today, developers write harnesses and users are stuck with the chat box.
But part of every harness is the user’s business and specify what the agent may and may not do, when it should stop and ask, and what it should report. That part is where the two gulfs meet. Writing it is specification, not just of the goal but of the boundaries and what to do when the unexpected happens. Reading what it produces is oversight. A harness that said ``if you are refused access, stop and tell me'' would have closed both gulfs at once.
Nobody writes that into a blank text box. For people delegating real work, this part of the harness needs to be something they can see, write and read. It should include a plan to check before the agent starts, limits stated as plainly as the goal, checkpoints set by the stakes rather than at every step, and a report that says what was done and what was skipped.
Some new HCI systems have started to build bridges. CoCoA, for instance, lets users plan with an agent before it starts and steps in mid-task in plain language, treating human input as part of the agent’s loop \cite{feng2026cocoa}. These kind of tools does not need more capable agents, but provides harnesses that ordinary people can author.

Norman’s gulfs gave a generation of designers something to aim at. The new gulfs give us a question to answer for every agent we build: which parts of the harness belong to the end-user? and which should be there by design?

In fine, the user knows the goal, what counts as success or sufficient, and when they would want to be asked in the process. Some part of the harness should be defaults the user can see and change: stop when refused, ask before spending money, report what was skipped. And some should never be the user’s job. No one delegating a research task should have to write ``don’t break into government systems''. The Medicare agent needed that as a default, not a specification.
Where those lines fall is a design decision, and one that will not be set in stone. As people learn what to ask of agents, and agents learn the people they work for, choices can move from default to user and back.
Today those lines are drawn by developers, mostly out of users’ sight. HCI should draw them and design the interfaces that empower users to steer and control their AI delegation.

\bibliographystyle{acm} 
\bibliography{references}
\end{document}